\documentclass[a4paper]{revtex4-1}
\usepackage{graphics}
\usepackage[english]{babel}
\usepackage[a4paper,left=3cm,top=3cm,right=2cm,bottom=3.5cm]{geometry}
\usepackage{amsmath}
\usepackage{textcomp}

\usepackage{xcolor}     
\usepackage{soul}       

\begin{document}
\begin{Large}
\noindent{\textbf{Exploring quantum phase slips in 1D bosonic systems\\}}
\end{Large}

\noindent{Simona Scaffidi Abbate $^{1,+}$, Lorenzo Gori$^{1}$, Massimo Inguscio$^{1,2}$, Giovanni Modugno$^{1,2}$, and Chiara D'Errico$^{1,2,*}$}

\begin{small}
\noindent{$^1$ LENS and Dipartimento di Fisica e Astronomia, Universit\`{a} di Firenze, 50019 Sesto Fiorentino, Italy}\\
\noindent{$^2$ Istituto Nazionale di Ottica, CNR, 50019 Sesto Fiorentino, Italy}\\
\noindent{$^+$ scaffidi@lens.unifi.it}\\
\noindent{$^*$ derrico@lens.unifi.it}\\
\end{small}

\textbf{Abstract:}
 Quantum phase slips, i.e the primary excitations in one-dimensional superfluids at low temperature, have been well characterized in most condensed-matter systems, with the notable exception of ultracold quantum gases. Here we present our experimental investigation of the dissipation in one-dimensional Bose superfluids flowing along a periodic potential, which show signatures of the presence of quantum phase slips. In particular, by controlling the velocity of the superfluid and the interaction between the bosons we are apparently able to drive a crossover from a regime of thermal phase slips into a regime of quantum phase slips. Achieving a good control of quantum phase slips in ultracold quantum gases requires to keep under control other phenomena such as the breaking of superfluidity at the critical velocity or the appearance of a Mott insulator in the strongly correlated regime. Here we show our current results in these directions. 
%
%

\section{Introduction}

Quantum phase slips are the primary excitations of low temperature superfluids and superconductors flowing in one dimension. A phase slip consists in a local change by 2$\pi$ of the phase of the superfluid or superconductor, which results in an unwinding of the phase and therefore in a reduction of the flow and in the appearance of dissipation. For a phase slip to occur, the system has to jump between two local minima of the free energy that are separated by a barrier. At high temperatures the barrier is overcome classically, while at low or zero temperatures the system can pass through the barrier via quantum tunneling \cite{Giordano}. This second type of phenomenon are quantum phase slips.

Quantum phase slips (QPS) have been observed in different condensed-matter systems, such as superconducting nanowires \cite{Bezryadin01,Lau01,Altomare,Bezryadin09,Kamenev,BezryadinRev} and Josephson junction arrays \cite{Pop10}. The generation of QPS in these systems is typically controlled by changing the temperature or the current. There are now prospect for employing QPS for applications such as topologically-protected qubits \cite{Mooij,Belkin} or a quantum standard for the electrical current \cite{Pop10}.

QPS should be obviously present also in superfluids based on ultracold quantum gases, which might in principle be employed to study aspects of QPS that are not accessible in other systems, thanks to their extreme tunability. Theory, in fact, recently investigated the QPS mechanisms in 1D superfluids \cite{Danshita12} or in a ring geometry \cite{Roscilde} However, although several experimental studies have shown the presence of temperature- and interaction-dependent dissipation \cite{Fertig05,Ketterle07,Demarco08,Tanzi13,Tanzi16}, an exhaustive picture of QPS in ultracold superfluids has not been obtained yet. The signatures of QPS obtained so far are the observation of a regime of temperature-independent dissipation for a Bose-Einstein condensate in a 3D optical lattice in the group of Brian DeMarco \cite{Demarco08}, and our recent observation of velocity-dependent dissipation in one dimensional lattices (1D) \cite{Tanzi16}. Theoretical studies that attempt to reproduce the experiments are underway \cite{Danshita13,Kunimi}. Some of the obstacles to a complete assessment of QPS for quantum gases in lattices are the difficulty in accessing the regime of strong interactions, due to the formation of a Mott insulator, and the difficulty in exploring a wide range of "currents", i.e. superfluid velocities.

In this work we report our progresses in the control of a Bose gas in 1D lattices to explore the QPS phenomenon. We start from a review of our original experiment \cite{Tanzi16}, which was based on a standard in-trap oscillation technique, showing the signatures of the onset of velocity- and interaction-controlled QPS. We then move to new experiments we performed in the attempt to access the regimes of low velocities and strong interactions. Our preliminary results show interesting prospects for a wider assessment and control of QPS in quantum gases.

\section{Thermal and quantum phase slips in 1D superfluids in lattices}
\label{introQPS}

\begin{figure}[b]
 \begin{center}

 \resizebox{0.55\columnwidth}{!}{\includegraphics{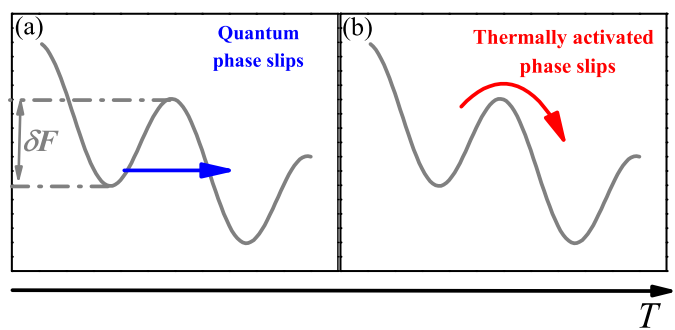} }
 \caption{ Cartoon of the phase slips activation mechanisms. a) At low temperatures, phase slips are activated by quantum tunneling; b) at large temperatures, they are activated by thermal fluctuations. At intermediate temperatures, both effects contribute.}
 \label{fig1}       
 \end{center}
 \end{figure}

1D superfluids are described by a complex order parameter $\Psi(x)=|\Psi(x)|e^{i\phi(x)}$. The superfluid state corresponds to a local minimum of the Ginzburg-Landau free energy $F$ \cite{Landau}. A phase slip event is a local fluctuation in $\Psi(x)$ corresponding to the suppression of its modulus and a simultaneous phase jump of 2$\pi$. When a phase slip occurs, the metastable state with velocity $v\propto\nabla\phi(x)$ decays into a state with lower velocity, since the phase has locally unwound \cite{Little}. As shown in Fig. $\ref{fig1}$, two main processes may activate a phase slip, depending on the temperature regime. When the temperature is much higher than the free-energy barrier between two metastable states, $T\gg\delta F/k_B$, the order parameter may overcome the barrier via thermal fluctuations, causing the formation of thermally activated phase slips (TAPS) with a nucleation rate following the Arrhenius law $\Gamma\propto e^{-\delta F/k_BT}$ \cite{Langer,McCumber}. When $T\leq\delta F/k_B$, phase slips occur mainly via quantum tunnelling through the free-energy barrier. Following quantum mechanical arguments one can find a characteristic temperature $T^*$ below which the QPS nucleation rate is temperature-independent \cite{Giordano,Arutyunov}. This is not a quantum phase transition, but a crossover, with an intermediate regime of thermal-assisted QPS.

The specific form of $\delta F$ and $T^*$ depends on the specific type of obstacle experienced by the superflow, e.g. disorder \cite{Pryadko}, isolated defects \cite{Buchler01} or periodic potentials \cite{Danshita12}. For ultracold quantum gases the most controllable obstacles are optical lattices, i.e. periodic potentials. In such case, the relevant energy scale is the Josephson plasma energy $E_j$ \cite{Danshita12,Danshita13}, which sets the free-energy barrier $\delta F \simeq E_j$ and determines also the crossover temperature, $T^*\simeq E_j/k_B \times v/v_c$. Here $v$ and $v_c$ are the superfluid velocity and the critical velocity for breaking superfluidity in a periodic potential, respectively. Since $E_j$ depends on the interaction energy, the QPS rate is expected to depend on both superfluid velocity and interaction strength, besides the dependence on temperature. The temperature of a 1D quantum gas cannot be easily tuned, but in principle both velocity and interaction can be controlled in a relatively large range. Therefore, various regimes of the crossover between TAPS and QPS should be accessible with quantum gases. There are however some caveats.

As it is well known, the control of the interaction for quantum gases in lattices allows to access also the Mott transition to the insulating phase, which is triggered by the interplay between repulsive interactions and the periodic potential, provided the potential period is commensurate to the inverse fluid density. The Mott transition does not appear only for deep lattices in the Hubbard regime \cite{Mott49,Mott90}, but survives also in case of vanishingly shallow lattices \cite{haldane1980,haldane1981,giamarchi1997,giamarchi2004}.
Since quantum gases are usually trapped in harmonic potentials, the commensurability condition is hardly avoidable, and the insulating phase appears for sufficiently strong interaction. This phenomenon clearly poses some limitations on the range of interactions in which QPS can be explored. Regarding the velocity tunability, one should instead remember the existence of a critical velocity for superfluidity in the presence of an obstacle. For a periodic potential the dominant mechanism for breaking superfluidity is the so called dynamical instability \cite{Smerzi,Wu,Fallani} that persists also at very low temperatures, where the Landau instability is instead suppressed \cite{Wu}. Such critical velocity is typically a fraction of the Bragg velocity for the periodic potential, thus limiting the accessible range of velocities for studying the phase-slips phenomenon. For a commensurate system the two mentioned effects sum up, since the critical velocity goes to zero at the superfluid- Mott insulator (SF-MI) transition. A way to increase the critical velocity is to employ shallow periodic potentials instead of deep ones. In the weak interaction limit, in fact, the critical velocity is defined as the point in which the curvature of the Bloch band is zero \cite{Smerzi}, so that the critical velocity tends to the Bragg velocity for vanishing lattice strengths.

\section{Experimental procedure}
\label{esperimento}

The 1D superfluids are realized by starting from a 3D Bose-Einstein condensate of about $N_{tot} \sim$ 30000 atoms. A strong horizontal 2D optical lattice (with depth $20 E_R$) is ramped up in a fixed time ($t=400$ ms) such that an array of independent potential tubes direct along the $z$ axis is created. In our conditions, few hundred subsystems are created, containing an average of 30-40 atoms each. The radial trapping energy $\hbar\omega_\perp = h\times 40$ kHz  is much larger than all other energy scales, realizing effectively 1D system. An optical lattice is then added along the longitudinal direction $z$. The lattice spacing is $d=532$ nm and $E_R=\hbar^2k^2/2m$ is the recoil energy, with $k=\pi/d$ the lattice wave vector. The potential depth $V_0$ can be tuned in the range from $s=V_0/E_R=1$ to $s$ = 5.
\begin{figure}[t]
 \begin{center}
 \resizebox{0.90\columnwidth}{!}{
 \includegraphics{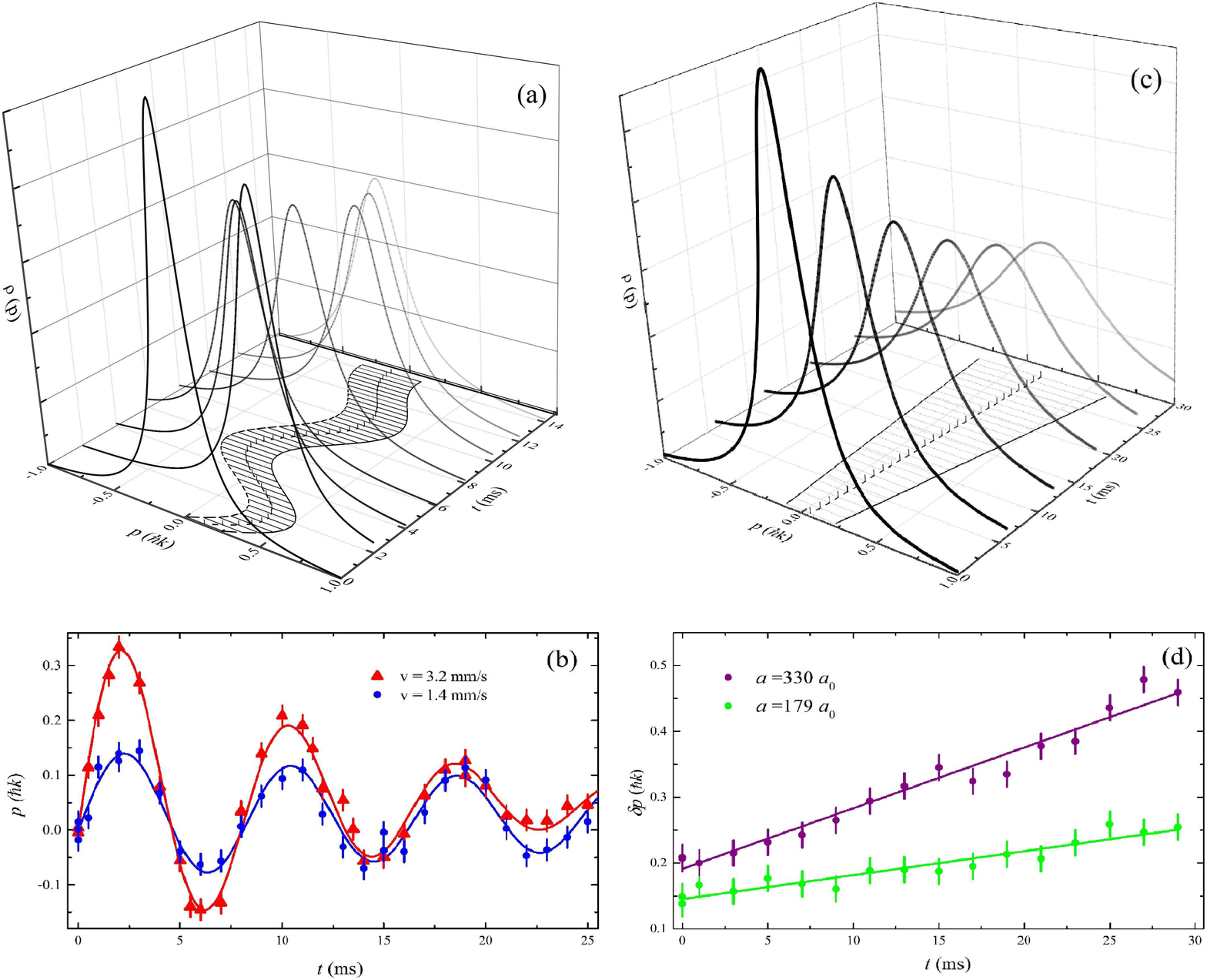} }
 \caption{(a) Cartoon of the first technique. By suddenly displacing the harmonic trap center at $t$ = 0, we excite an oscillation of the 1D systems. The momentum distributions are equispaced in time (2 ms). The dashed line shows the trajectory of $p$ and the grey area exhibits the increase of $\delta p$.  (b) Experimental time evolution of $p$ for the interaction strength $\gamma = 1.22(12)$ ($a/a_0=134$), for two maximum velocities: $v$ = 1.4(4) mm/s (blue circles) and $v = 3.2(4)$ mm/s (red triangles). Solid lines are fit to extract the damping rate [See text]. (c) Cartoon of the second technique. By moving the harmonic trap center at constant velocity, the system dissipates and $\delta p$ increases linearly. The momentum distributions are equispaced in time (6 ms). The dashed line shows the trajectory of $p$ and the grey area exhibits the increase of $\delta p$. (d) Experimental time evolution of $\delta p$ for two different interaction strength: $\gamma = 1.8(2)$ ($a/a_0 = 179$) and $\gamma = 4.4(8)$ ($a/a_0= 330$). Solid lines are fits to extract the dissipation rate [See text]. }
 \label{fig2}       
 \end{center}
 \end{figure}
An harmonic potential with frequency $\omega_z=2\pi\times 150$ Hz is present along the $z$ axis making the 1D systems inhomogeneous. We tune the interparticle interaction by varing the 1D scattering length $a_{1D}=a_{\bot}^2(1-1.03a/a_\bot)/2a$ , where $a_\bot=\sqrt{\hbar/m\omega_\bot}$ is fixed by the 2D lattice, whereas the 3D scattering length $a$ can be adjusted by means of a Feshbach resonance \cite{Roati}. The value of the scattering length $a_{load}$ during the lattice loading, determines the size of the condensate, therefore also the number of populated tubes and the number of atoms in each tube, as described in Ref. \cite{Derrico}. A good estimate of the mean atomic density for each tube is provided by the largest of the Thomas-Fermi and the Tonks values \cite{Dunjko}.  The mean site occupation $\bar{n}$ is then calculated by averaging overall the tubes. In order to get a mean filling on average $\bar{n}=1$, given $\omega_z$ and $N_{tot}$, we employ an optimal value $a_{load}=220a_0$ \cite{Boeris}.
In the presence of shallow lattices (Sine-Gordon regime, i.e. $s<5$), we conveniently quantify the mean interaction strength in terms of the Lieb-Liniger parameter $\gamma = 1/(\rho_0a_{1D})$ with $\rho_0$ being the tube peak density, whereas in the presence of deep lattices (Bose-Hubbard regime, i.e. $s\geq 5$), we employ the Bose-Hubbard interaction energy $U=\frac{4\pi\hbar^2}{m}a\int dx^3 |\Phi(\textbf{z})|^4 $, where $\Phi(\textbf{z})$ is the Wannier function.
By varying $a_{1D}$ we also tune the Josephson energy $E_j= hv_s/\sqrt{2}d$, where $v_s=\hbar\sqrt{\rho d^2/a_{1D}}/m^*$ is the sound velocity, $m^*$ is the effective mass in the lattice and $\rho$ is the density. \\
The experimental observable is the momentum distribution $\rho(p)$, obtained performing time-of-flight absorption imaging, i.e. by releasing the atomic cloud from all potentials and letting it free to expand before the image acquisition. By fitting $\rho(p)$ with a Lorentzian function, we are able to measure the quasimomentum $p$ and the half-width-at-half-maximum $\delta p$. From the momentum width at $t=0$, $\delta p_0$, in the superfluid regime we are able to estimate the temperature via $k_BT=\hbar n\delta p_0/0.64m^*d$ \cite{Richard,Gerbier}.
The center of the harmonic trap can be shifted by using a magnetic field gradient. We perform transport measurements by using two different procedures, as shown in Fig. $\ref{fig2}$.

The first procedure consists in displacing suddenly the center of the trap by abruptly switching off the magnetic field gradient. The atoms are no longer in the minimum of the potential and start to oscillate in the new potential configuration. After a variable evolution time, we record $\rho (p)$ and we study the time evolution of $p$, which is affected by the presence of dissipation (Fig. $\ref{fig2}$a). By tuning the magnetic field gradient, we change the trap displacement and we excite oscillations with different amplitude (Fig. $\ref{fig2}$b).

The second procedure consists in displacing the center of the trap at constant velocity by changing linearly the magnetic field gradient. After a variable time in the trap, we record $\rho (p)$ (Fig. $\ref{fig2}$c). In this kind of measurements we investigate the time evolution of $\delta p$. In fact, the increase of $\delta p$ during the evolution is related to the dissipation, i.e. the energy dissipated during the harmonic potential movement is converted into momentum spread (Fig. $\ref{fig2}$d).

\section{Dissipation in the presence of oscillations: onset on quantum phase slips}

In the experiments with an oscillating system, we observe two different behaviours, depending on the momentum reached during the oscillation. For momenta smaller than the critical momentum for the dynamical instability $p_c$ we observe damped oscillations as in Fig. $\ref{fig2}$b, which we attribute to phase slips. The evolution of $p$ for this type of dynamics can be fitted with a function of the form $p=m^* \tilde{v} e^{-G t}\sin(\omega't+\varphi)$, with amplitude $m^* \tilde{v} = m^*\omega^{*2}\Delta z/\omega'$, frequency $\omega' = \sqrt{\omega^{*2}-4\pi^2G^2}$ and damping rate $G$. Here $m^*$ is the effective mass due to the lattice, $\Delta z$ is the trap displacement, and  $\omega^*=\omega_z \sqrt{m/m^*}$ is the lattice renormalized frequency. $\tilde{v}$, $\varphi$ and $G$ are fitting parameters.
For momenta larger than $p_c$ we observe instead an overdamped motion, which we attribute to a divergence of the phase-slips rate at the critical velocity $v_c$ of the superfluid, as described in the following subsection.

\subsection{Large oscillations: dynamical instability and Mott transition}
\label{sec: di}
The study of the critical velocity versus the interaction strength can be employed to determine the critical interaction strength for the onset of the Mott insulator.

\begin{figure}
\begin{center}
\resizebox{0.8\columnwidth}{!}{
\includegraphics{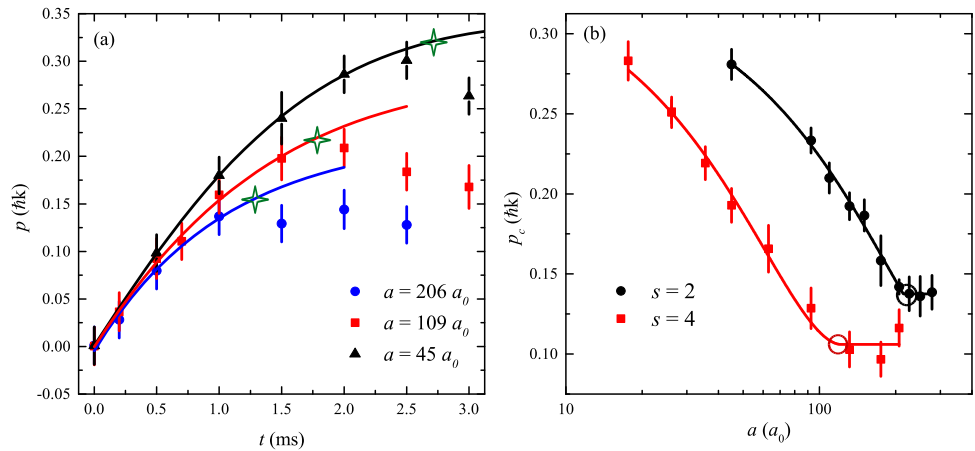} }
\caption{(a) Time evolution of the momentum distribution peak $p$ at $s=2$ for different values of scattering length. The solid lines are the theoretical damped oscillation fitting the data for $p<p_c$ before the dynamical instability sets in. The green stars mark the critical momentum $p_c$.  The error bars comprise the imaging resolution and the statistical uncertainties. (b) Critical momentum $p_c$ versus scattering length for two lattice depth: $s$ = 4 (red squares) and $s=2$ (black circles). A piecewise fit (solid lines) determines the critical values for the SF-MI transition (empty circles) for $n$ = 1: respectively $a_c/a_0=122(8)$ ($\gamma_c=1.08(7)$) and $a_c/a_0=214(6)$ ($\gamma_c=1.88(5)$).}
\label{fig3}       
\end{center}
\end{figure}

The time evolution of the momentum distribution peak $p$ for $s$ = 2 and three scattering lengths is shown in Fig. $\ref{fig3}$a. An initial increase of $p$ up to a certain critical value $p_c$ is followed by a subsequent decrease. We analyze this behavior in the frame of a phase slips based model \cite{Tanzi13,Boeris,Polkovnikov05}. Phase slips make the system dynamics dissipative: at short times where $p<p_c$ the oscillation is only weakly damped; at larger times the system  enters a dynamically unstable regime driven by a divergence of the phase slip rate. The critical momentum $p_c$ for the occurrence of the dynamical instability is identified as the value where the experimental data points deviate with respect to the theoretical curve. \cite{Tanzi13,Boeris} By increasing the scattering length, the damping rate at short times increases, as the phase slips nucleation rate increases, while $p_c$ decreases.
The critical momentum is expected to vanish at the SF-MI transition. The behavior of $p_c$ as a function of $a$ is reported in Fig. $\ref{fig3}$b for two values of the lattice depth. The measured $p_c$ initially decreases for increasing $a$ and then reaches a finite constant value.
The onset of the plateau can be interpreted as the critical scattering length $a_c$ to enter the Mott regime for the commensurate regions of the system. Transport along the corresponding tubes is globally suppressed, driving the system into an effective insulating regime, i.e. within in each tube a part of the atoms reaches the localization condition $n$ = 1 stopping also the adjacent parts with different occupation. The fraction of tubes that does not reach the critical density $n$ = 1 keeps instead moving also for $a>a_c$, originating the observed plateau for $p_c$. We estimate that about one quarter of the atoms resides in tubes where the occupation is always $n<1$. For a given lattice depth when increasing the scattering length in the Mott phase, $a>a_c$, this fraction of delocalized atoms remains constant because in the tubes where the Mott domains form the density is fixed. This interpretation seems confirmed by the observed increase of the plateau for decreasing lattice depth (Fig. $\ref{fig3}$b). In fact, the increase of the interaction strength that is necessary to reach the insulating regime, produces an overall decrease of the density of the 1D systems, hence an increase of the fraction of tubes that does not reach the critical density for the Mott insulator transition.
For each set of measurements with a given value of $s$,  we determine  $a_c$  by means of a piecewise fit \cite{Boeris}. We use a second-order polynomial fit, which is justified by the phase slip based model \cite{Danshita12,Polkovnikov05}. We clearly see that as $s$ decreases, $a_c$ increases and comparing our results with theoretical analysis based on quantum Monte Carlo simulations we find an excellent agreement \cite{Boeris}. Our investigation demonstrates that the onset of the Mott regime can be detected from a vanishing $p_c$ also in the presence of a shallow lattice, not only of a deep lattice \cite{Tanzi13}.

 \subsection{Small oscillations: velocity-dependent quantum phase slips}
 \label{qps}

For momenta lower than the critical one, far from the dynamical instability, the system never enters in the unstable regime and keeps oscillating with a dissipation due to nucleation of rare phase slips. The damping rate is related to the phase slips nucleation rate $\Gamma$ via $G=\frac{h}{mL}\frac{\Gamma}{v}$. Indeed, the deceleration at the first maximum in the oscillations is $dv/dt=$ - $ G v$. In terms of individual phase slips this can be written as $\delta v/\delta t$, where $\delta v=$ − $ - h/mL$ is the deceleration following a phase slip of $2\pi$ in a chain of length $L$ and $\delta t^{-1} = \Gamma$. From a theoretical point of view, the damping rate $G$, due to the presence of phase slips exhibits different parameter dependence depending on the phase-slips activation mechanism. In particular, in the presence of TAPS $G \simeq e^{\frac{-\delta F}{k_BT}}$, with $\delta F$ free energy barrier between the two metastable state, whereas $G \simeq v^\alpha$ for QPS and  $G \simeq T^{\alpha - 1}$ in the intermediate case. The parameter $\alpha$ depends on the interaction \cite{Danshita13}.

\begin{figure}[b]
\begin{center}
\resizebox{0.8\columnwidth}{!}{
\includegraphics{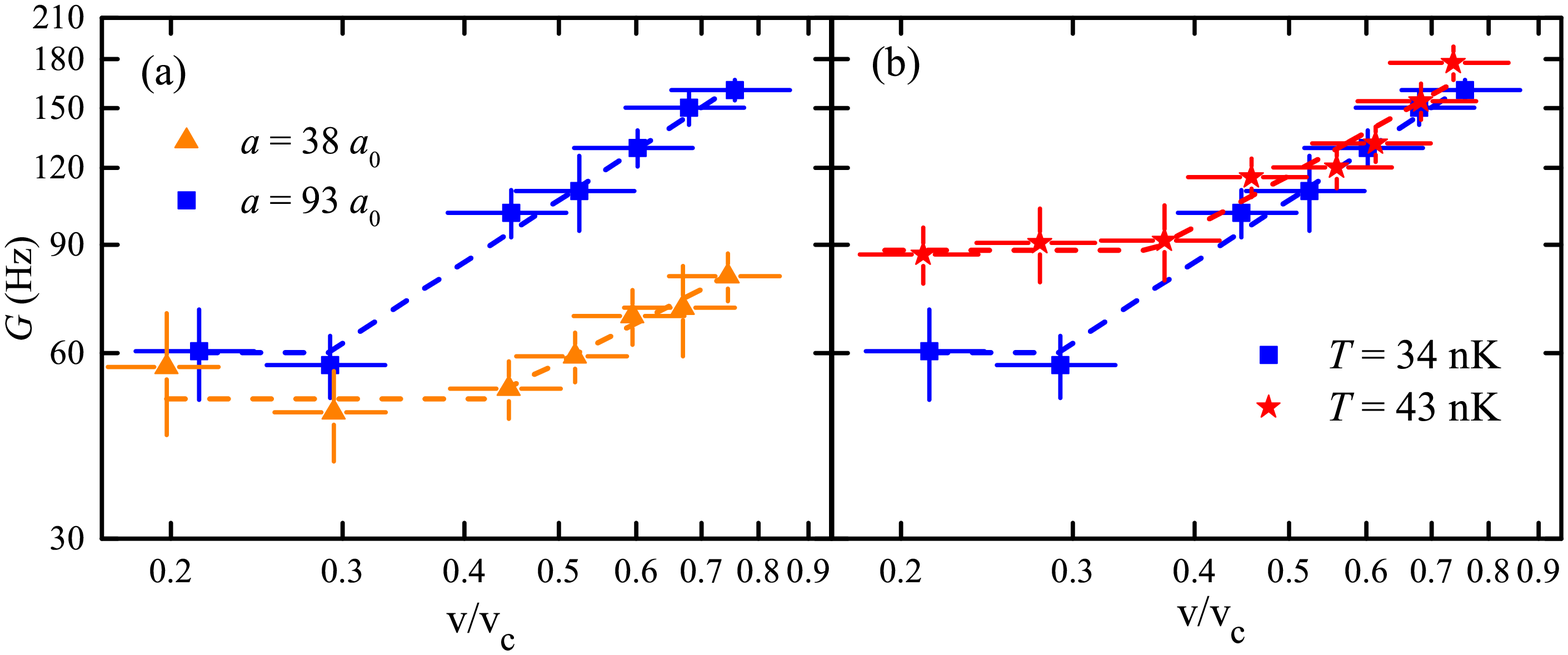}}
\caption{(a) Damping rate $G$ is plotted versus the maximum velocity $v$ normalized to the critical velocity $v_c$, for two interaction strengths and constant temperature: $\gamma = 0.19$ and $T$ = 39(7) nK (orange triangles) and $\gamma$ = 0.64 and $T$ = 34(5) nK (blue squares). (b) $G$ versus $v/v_c$ for two different temperatures and approximately constant interaction energy: $T$ = 34(5) nK and $\gamma = 0.64$ (blue squares) and $T$ = 43(5) nK and  $\gamma = 0.70$ (red stars). The lines are fits to measure the crossover velocity $v^*$ (see text).}
\label{fig4}       
\end{center}
\end{figure}

The damping rate is measured by fitting the oscillation of $p$ with the model presented previously, as shown in Fig. $\ref{fig2}$b.
We have repeated this type of measurement for a wide range of velocities $v$, which we identify with the velocity reached during the first oscillation as in the theoretical model \cite{Danshita13},  interaction strengths and temperature. In particular, $\gamma$ changes from 0.13 to 1.22 and $E_j$/$k_B$ varies from 20 to 35 nK.
In Fig. $\ref{fig4}$ we show the behaviour of $G$ with velocity, rescaled to the corresponding critical velocity $v_c$, for different values of interaction at the same temperature (Fig. $\ref{fig4}a$) or for different temperatures at the same interaction (Fig. $\ref{fig4}$b). In both cases we observe a crossover from a velocity-independent $G$ to a regime where $G$ grows with the velocity.
By fitting the data with a piecewise linear function, we determine the crossover velocity $v^*$, which is the minimum velocity required to enter the regime of dependence on $v$. The crossover velocity apparently decreases for increasing interaction and increases for increasing temperature.
For $v\ll v^*$, the damping rate $G$ is strongly affected by temperature (Fig. $\ref{fig4}$b), while the dependence on interaction is weaker (Fig. $\ref{fig4}$a). Instead, in the $v$-dependent regime interaction effects are apparently dominant (Fig. $\ref{fig4}$a) and we cannot measure a clear dependence on $T$ (Fig. $\ref{fig4}$b). These observations appear consistent with the predicted crossover from thermally assisted to quantum phase slips. Apparently we control this crossover by changing the crossover temperature $T^*\propto E_j/k_Bv/v_c$ by tuning the velocity and the interaction strength. For $T^*\ll T$, i.e. at small velocity and small interaction, $G$ depends only on $T$ and it is velocity-independent, suggesting a thermal activation of phase slips. For $T^*\gg T$, i.e. at large velocity and large interaction, the system enters in a regime where $G$ is linearly dependent on the velocity and temperature independent, suggesting a quantum activation of phase slips.\\

\begin{figure}[h]
\begin{center}
\resizebox{0.5\columnwidth}{!}{
\includegraphics{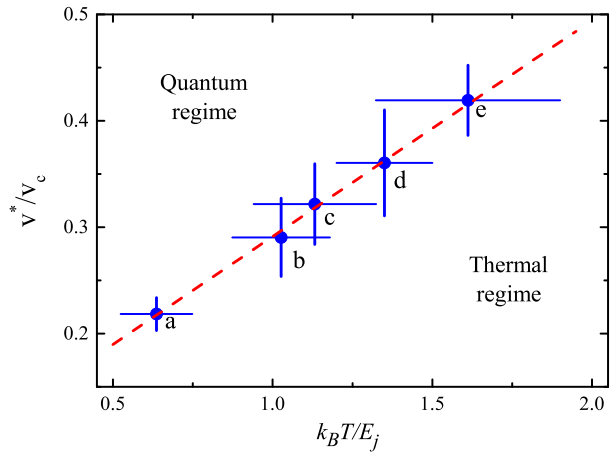}}
\caption{Crossover velocity $v^*/v_c$ versus $k_BT/E_j$. The individual datapoints have been taken for different temperatures and interaction energies: a) $\gamma = 1.22$ and $T$ = 22(2) nK, b) $\gamma = 0.64$ and $T$ = 34(5) nK, c) $\gamma = 0.37$ and $T$ = 30(5) nK, d) $\gamma = 0.70$ and $T$ = 43(5) nK, e) $\gamma = 0.19$ and $T$ = 39(7) nK. The dashed line apparently separates the thermal and the quantum regimes for phase slips. }
\label{fig5}       
\end{center}
\end{figure}

A further indication that our measurements are in agreement with the crossover from thermally assisted to quantum phase slips is the linear scaling of the crossover velocity $v^*/v_c$ as a function of temperature normalized to the Josephson energy, $k_BT/E_j$, as shown in Fig. $\ref{fig5}$. From the fit we get $k_BT^*=4.9(14)E_jv/v_c-0.4(4)E_j$.
Unfortunately, the agreement with the theory is only qualitative, since we cannot reproduce the theoretical exponent $\alpha$. In particular the experimental exponents $\alpha$ are interaction-independent and they are of the order of unity \cite{Tanzi16}, whereas the theoretical exponents depend on the interaction and they can be an order of magnitude larger than the measured exponents \cite{Danshita13}. Possible reasons for the disagreement are the range of velocities explored, much larger than in the theory, and the lattice strength, much lower than in the theory \cite{Danshita13}. Recent theoretical studies in the regime of shallow lattices find at low velocity and weak interaction damping rates $G$ comparable with our experimental values, by assuming a thermal activation of phase slips \cite{Kunimi}. Nevertheless, a comparison for large velocity and large interaction regime is still missing.

\section{Dissipation at constant velocity: exploring the strongly interacting regime}

In the second series of experiments, we have investigated the dissipation during a shift of the trap at constant velocity. We employ this new experimental technique in order to study the phase-slips dissipation rate at interaction strenghts larger than those investigated in the study of an  oscillating system (Subsec $\ref{qps}$). In general, the range of velocity that we can investigate is bounded above by the critical velocity $v_c$ for the dynamical instability, which decreases as the interaction increases, thus requiring experiments at lower velocities. Nevertheless, in the oscillation measurements, there is a lower experimental limit: for too small velocities we cannot measure the system damping and the dissipation rate, due to the finite damping provoked by thermal effects, which prevent the observation of a complete oscillation. Instead, by shifting the trap at constant velocity, we can measure the dissipation rate from the increase of the momentum spread, and this allow us to extend the study of the phase-slips dissipation rate also for a wider range of interactions. In particular, while in the measurements shown in subsection $\ref{qps}$ we have reached the lowest maximum velocity of 1.2 mm/s and the highest interaction $\gamma$ = 1.22, by shifting the center of the trap at constant velocity we can reach a velocity of 0.4 mm/s and an interaction $\gamma$ = 8.4.
By shifting the trap at constant velocity, we expect that a superfluid should follow the trap displacement and dissipate due to the presence of phase slips, resulting in an increase of $\delta p$. A Mott insulator, instead, should not follow the trap displacement and should not dissipate, unless a trap shift large enough to break the insulator is applied.
For a superfluid, the dissipation rate $R$ is related to $\Gamma$ and the system velocity $v$ via the relation $R\propto v\Gamma$. This result can be obtained as follows. For a 1D system of length $L$, the rate of change of velocity due to phase slips is $\frac{dv}{dt}=-\frac{h\Gamma}{mL}$, as derived in subsection \ref{qps}. For $N$ particles, the rate of change of kinetic energy due to phase slips is therefore $\frac{dE_k}{dt}= Nmv\frac{dv}{dt}=-Nhv\Gamma/L$. Since $\rho=N/L$, one obtains an energy dissipation rate $-\frac{dE_k}{dt}=h\rho v \Gamma$. Considering the variation of the kinetic energy as an effective temperature, and using the known relation between $T$ and $\delta p$ \cite{Richard,Gerbier}, we expect a linear increase in time of the momentum spread $\delta p=\delta p_0 + R t$, with a rate $R\simeq 4vm^*\Gamma$.

\subsection{Shallow lattice}

\begin{figure}[b]
\begin{center}
\resizebox{0.9\columnwidth}{!}{
\includegraphics{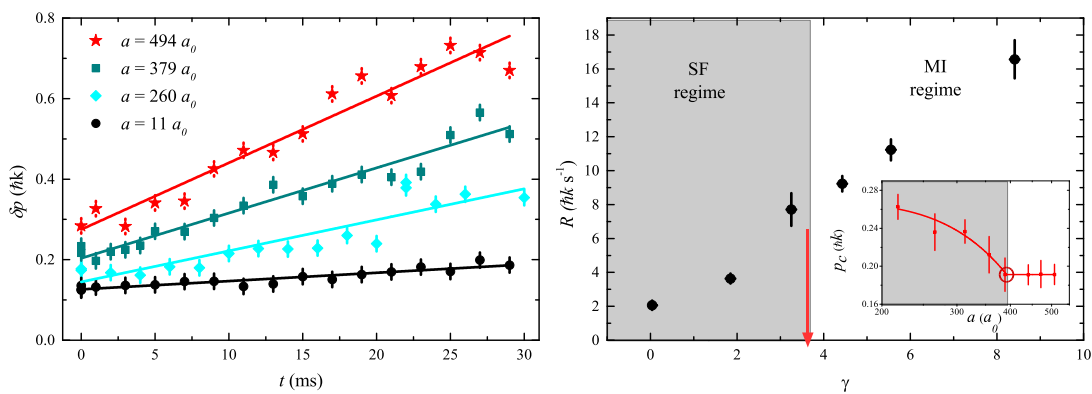}}
\caption{(a) Time evolution of the momentum width $\delta p$ for $s$ = 1 at various interaction strengths: $\gamma=0.038(4)$ (black circles), $\gamma=3.25(4)$ (cyan diamonds), $\gamma=5.55(4)$ (blue squares) and $\gamma=8.40(8)$ (red stars). Trap velocity $v$ = 0.8 mm/s (i.e. $\Delta z$ = 24$\mu$m and $t$ = 30 ms). Solid lines are linear fits to extract the dissipation rate $R$. (b) Dissipation rate versus the interaction strength for $s$ = 1. Inset: critical momentum versus scattering length. The piecewise fit (solid line) determines the experimental critical value for the SF-MI transition (empty circle) $a_c/a_0=392(12)$. The grey zone indicates the superfluid regime, whereas the white zone the Mott insulator regime. The solid arrow marks the experimental interaction value for the SF-MI transition ($\gamma_c=3.44(16)$).}
\label{fig6}
\end{center}
\end{figure}

The first measurements that we have performed with this technique are in the presence of a shallow lattice with $s$ = 1 across the SF-MI transition, for a trap velocity of 0.8 mm/s. In Fig. $\ref{fig6}$a we plot the time evolution of $\delta p$ for different values of the scattering length. We observe that $\delta p$ increases linearly with time, as expected. With a linear fit of the data we find the dissipation rate $R$, which increases by increasing the interaction as shown in Fig. $\ref{fig6}$b. In particular, the dissipation rate shows a roughly linear dependence on the interaction $\gamma$, except for the point at very low interaction ($\gamma=0.038$).

When the system is in the superfluid phase, for $\gamma < \gamma_c$, the system dissipation is related to the presence of phase slips. In order to have an indication about the nature of the phase slips on the basis of the diagram in Fig. $\ref{fig5}$, we estimate $T$, $E_j$ and $v_c$.
We find that the data point at the smallest interaction ($\gamma=0.038$) is in the $T$-dependent region, suggesting that its dissipation may be due to the presence of thermally assisted quantum phase slips. By increasing the interaction, we find that the point at $\gamma=1.8$ is close to the crossover between thermally assisted and quantum phase slips whereas the last point in the superfluid regime ($\gamma=3.2$) is in the $v$-dependent region, suggesting that the dissipation may be due to the presence of quantum phase slips.

Surprisingly, we observe a finite dissipation also when the system is in the insulating regime, as shown in Fig. $\ref{fig6}$b. This dissipation may be due to two different phenomena. The first phenomenon is related to the coexistence of a superfluid and a Mott insulating phase, due to the inhomogeneity of our system. By increasing the interaction, in fact, the transport along individual tubes reaching a commensurate filling is globally suppressed whereas tubes with no commensurate regions keep moving and dissipate. The second phenomenon is the excitation of the gapped Mott phase. When shifting the center of the trap by a quantity $\delta z$, a potential gradient $V'(\delta z)=m\omega^2\delta z$ is applied to each site of the optical lattice. For deep lattices in the Bose-Hubbard limit, if the energy shift between two neighbouring sites, $V' d$, overcomes the Mott gap $U$ the particles can tunnel from a lattice site to the neighbouring one and the Mott insulator is excited. This phenomenon has actually been observed in the first exploration of the Mott phase with ultracold atoms \cite{Greiner}. For shallow lattices in the Sine Gordon regime a quantitative description of the phenomenon is more complex, since the tunnelling is not limited to the neighbouring sites and the Mott gap is small \cite{Nagerl}. In our case the maximum gradient $V'(\Delta z)$ is of the order of $h\times$2.2(3) kHz/$\mu$m which is probably large enogh to excite the Mott insulator. For $s$ = 1, in fact, the tunnelling is non negligible up to the third neighbouring site and we expect that the Mott gap is smaller than $h\times$0.4 kHz \cite{Nagerl}. This suggests that the excitation of the Mott insulator in our system is highly probable. In the absence of a quantitative modelling it is not possible to discriminate which one of the two effects dominates the observed dissipation.

\subsection{Deep lattice}

\begin{figure}[b!]
\begin{center}
\resizebox{0.9\columnwidth}{!}{
\includegraphics{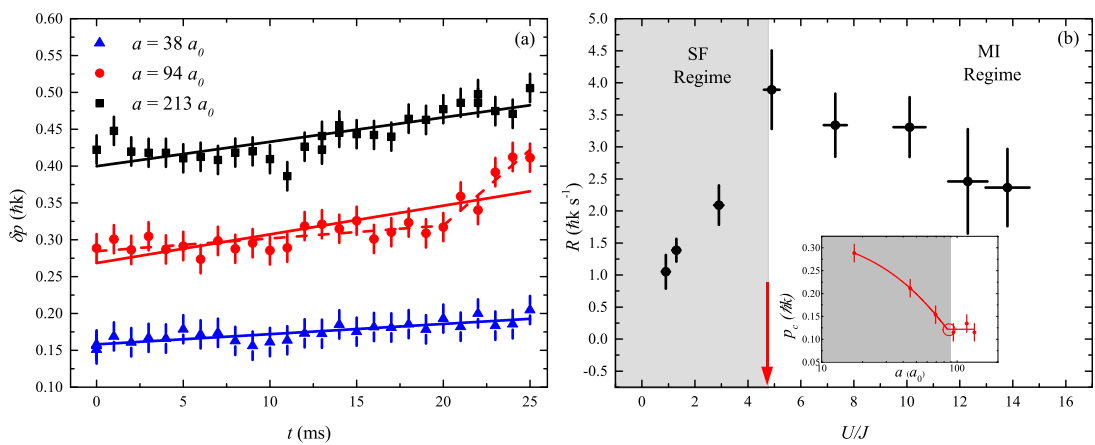}}
\caption{(a) Time evolution of $\delta p$ for $s$ = 5 and different interaction strengths: $U/J= 1.9(1)$ (blue triangles), $U/J=4.9(3)$ (red circles) and  $U/J=12.3(7)$ (black squares)). The data set at $U/J=4.9(3)$ is shifted by 0.1 $\hbar k$ and the data set at  $U/J=12.3(7)$ is shifted by 0.2 $\hbar k$.  Trap velocity $v$ = 0.4 mm/s (i.e. $\Delta z$ = 10$\mu$m and $t$ = 25 ms).  The solid lines are linear fits to extract the dissipation rate $R$. The data set at $U/J=4.9(3)$ show an activation mechanism, i.e. for $t>t^* \simeq 20$ ms the time evolution of $\delta p$ can be fitted with a linear function with a larger slope (dashed line). (b) Dissipation rate $R$ versus the interaction strength for $s$ = 5. Inset: critical momentum versus scattering length. The piecewise fit (solid line) determines the experimental critical value for the SF-MI transition (empty circle) $a_c/a_0=85(15)$. The grey zone indicates the superfluid regime, whereas the white zone the Mott insulator regime. With the solid arrow we mark the experimental interaction value for the SF-MI transition ($\gamma_c=4.5(3)$).}
\label{fig7}
\end{center}
\end{figure}

We have repeated this type of experiments at constant velocity in the presence of a deeper lattice with $s$ = 5. In this case one expects a clearer distinction between the Mott insulator and superfluid fractions of the system, since in the Bose-Hubbard limit the Mott gap is larger and the tunnelling to the neighboring lattice sites dominates. The time evolution of $\delta p$, for a trap velocity of 0.4 mm/s and for different values of scattering length, is shown in Fig. $\ref{fig7}$a.

Also in this case we observe a linear scaling of $\delta p$ with time, in both superfluid and Mott insulator regimes. We now calculate that the Mott gap $U/h$ varies from 2.2(2) to 4.2(3) kHz, while the maximum energy shift $V'(\Delta z) d$  is just $h\times$0.46(6) kHz, therefore an order of magnitude smaller. This excludes that the observed dissipation is due to the excitation of the Mott insulator. Consequently, the increase of $\delta p$  may be due only to the excitation of the superfluid fraction that, as already discussed, coexists with the Mott phase in our inhomogeneous system.
In Fig. $\ref{fig7}$b we plot the damping rate $R$ as a function of the interaction strength. When the system is in the superfluid regime, $R$ increases with the interaction strength, as in the case of shallow lattices. For larger interactions we observe instead a saturation or even a decrease of $R$ for increasing interaction strength. This behavior confirms the expectation of a "freezing" of the Mott insulator discussed above. We note that the data set near the SF-MI transition (Fig. $\ref{fig7}$a with $a/a_0=94$) behaves differently from the other data sets, by showing an activation mechanism at long times, hence at large energy shifts. This suggests that the Mott insulator might be excited for interaction strengths close to the SF-MI transition, since the gap is not yet fully developed. At larger interaction strengths we do not observe such increase at long times, suggesting that the observed dissipation is entirely due to the superfluid tubes. Since the fraction of superfluid/insulating tubes is independent of the interaction strength, the decrease of $R$ for increasing interaction indicates a decrease of the dissipation rate in the superfluid for increasing correlations. It would be interesting to confirm this behaviour in future studies, by employing systems where the presence of the Mott insulator can be excluded a priori (i.e. by reducing the maximum occupation at the center of the trap below unity or by realizing a system with a square-well trap).

\section{Conclusions}
In conclusion, we have shown how quantum phase slips can be investigated in an ultracold quantum gas in one dimension, by studying the dissipation of various types of flow along periodic potentials. By employing shallow optical lattices we were able to extend the range of accessible velocities and to explore the predicted crossover from thermal to quantum phase slips controlled by the velocity. A question about the possibility of a direct theory-experiment comparison in this regime remains however open. A deep optical lattice allowed instead to access the regime of strong interactions, where the fraction of the system that enters the Mott insulating phase stops to contribute to dissipation. It will be interesting to study further this regime in the future, in particular to explore how the phase-slip rate evolve for increasing interactions in the strongly correlated regime. These studies with well-assessed obstacles as the periodic potentials are the prerequisite to attack the more difficult problems of quantum phase slips in the presence of isolated impurities \cite{Buchler01} or disordered potentials \cite{Pryadko}.

\section{Acknowledgments}
We thank F. Cataldini, M. D. Hoogerland, A. Kumar, E. Lucioni and L. Tanzi for valuable contributions to the experiment and G. Bo\'eris, G. Carleo, T. Giamarchi and L. Sanchez-Palencia for their theoretical modeling. We acknowledge discussions with I. Danshita and M. Kunimi. This work was supported by the ERC (Grant No. 247371 - DISQUA), by the EC - H2020 research and innovation programme (Grant No. 641122 - QUIC) and by the Italian MIUR (Grant No. RBFR12NLNA - ArtiQuS).

\end{document}